\documentclass{emulateapj}

\def\gsim{\mathrel{\rlap{\lower 4pt \hbox{\hskip 1pt $\sim$}}\raise 1pt
\hbox {$>$}}}
\def\lsim{\mathrel{\rlap{\lower 4pt \hbox{\hskip 1pt $\sim$}}\raise 1pt
\hbox {$<$}}}

\begin{document}

\title{Injection and Acceleration of Electrons at A Strong Shock: \\
Radio and X-ray Study of Young Supernova 2011dh} 

\author{
Keiichi Maeda\altaffilmark{1}
}

\altaffiltext{1}{Kavli Institute for the Physics and Mathematics of the 
Universe (Kavli-IPMU), University of Tokyo, 
5-1-5 Kashiwanoha, Kashiwa, Chiba 277-8583, Japan; 
keiichi.maeda@ipmu.jp .}

\begin{abstract}
In this paper, we develop a model for the radio and X-ray emissions from Type IIb Supernova (SN IIb) 2011dh in the first 100 days after the explosion, and investigate a spectrum of relativistic electrons accelerated at a strong shock wave. The widely-accepted theory of the particle acceleration, so-called diffusive shock acceleration (DSA) or {\em Fermi} mechanism, requires seed electrons with modest energy with $\gamma \sim 1 - 100$, and little is known about this pre-acceleration mechanism: We derive the energy distribution of relativistic electrons in this pre-accelerated energy regime. We find that the efficiency of the electron acceleration must be low, i.e., $\epsilon_{\rm e} \lsim 10^{-2}$ as compared to the conventionally assumed value of $\epsilon_{\rm e} \sim 0.1$. Furthermore, independently from the low value of $\epsilon_{\rm e}$, we find that the X-ray luminosity cannot be attributed to any emission mechanisms suggested so far {\em as long as these electrons follow the conventionally-assumed single power-law distribution.}  A consistent view between the radio and X-ray can only be obtained if the pre-acceleration injection spectrum peaks at $\gamma \sim 20-30$ and then only a fraction of these electrons eventually experience the DSA-like acceleration toward the higher energy -- then the radio and X-ray properties are explained through the synchrotron and inverse Compton mechanisms, respectively. Our findings support the idea that the pre-acceleration of the electrons is coupled with the generation/amplification of the magnetic field. 
\end{abstract}


\section{Introduction}

Identifying the site of the particle acceleration, and the mechanism responsible for it, has been an intensively studied field of particle astrophysics. The most promising mechanisms require the existence of a strong shock wave, exemplified by the diffusive shock acceleration (DSA) mechanism (or {\em Fermi} mechanism) where the particles acquire energy through repeated collisions between up- and down-streams of the shock wave \citep{fermi1949,blandford1978,bell1978}. Supernova remnants (SNRs) are believed to be the origin of a major part of cosmic rays up to $\sim 10^{15}$eV \citep{baade1934,bamba2003,uchiyama2007}. The acceleration of hadrons there is still controversial \citep{ellison2000,ellison2007,morlino2012}, and studying the relativistic electron population helps clarify the issue since the essence of the acceleration must be the same, and understanding the electron population will enable us to subtract the electron contribution from the observed high energy emissions. 

There is one key open issue in this picture -- how the electrons are `pre-accelerated'. For the DSA mechanism works effectively, a particle must have an enough kinetic energy exceeding a thermal energy behind the shock, in order to cross the shock wave and experience the repeated collisions. The nature of this pre-acceleration mechanism should be seen in the energy distribution of non-thermal electrons in the MeV range, but so far this energy range is unexplored observationally in conventional astrophysical acceleration sites like SNRs. 

A young core-collapse supernova (CC SN) before entering the supernova remnant phase offers a unique opportunity here. The SN ejecta running into the circumstellar material (CSM) launches a shock wave \citep{chevalier1982}. At the shock wave a fraction of thermal particles are accelerated to the relativistic speed, and a magnetic field is generated/amplified. Relativistic electrons gyro around the magnetic field producing synchrotron emissions observed in radio \citep{weiler1989,chevalier1998}. X-rays are frequently detected from nearby CC SNe when the observations are performed with sufficient sensitivities \citep[see][and references therein]{immler2002}. For the X-ray emission mechanism, the emission from the thermal electrons behind the shock, the synchrotron emission from the relativistic electrons, and the inverse Compton (IC) scattering by the relativistic electrons reprocessing the optical photons from the SN ejecta to the X-ray energy, have been proposed \citep{chevalier2006b}. The physical condition is similar to SNRs, but there is one important difference in terms of the observational potential. The young SNe are characterized by the high magnetic field content at the shock front (due to the high CSM density). In SNRs, the typical (amplified) magnetic field strength ($B$) is $\sim 100 \mu$ Gauss \citep{bamba2003,uchiyama2007}. Thus, at $1$ GHz typical of radio observations, electrons with the Lorentz factor $\gamma \gsim 3000$ are responsible for such an emission. On the other hand, for a typical magnetic field $B \sim 1$ Gauss in young CC SNe \citep{chevalier2006b}, radio emission at $\sim 1$ GHz is produced by electrons with $\gamma \sim 100$. 

A detailed model of the electron acceleration and the non-thermal emissions requires an intensive set of observations in the radio and X-ray regime. It has been done so far for type IIb SN (SN IIb) 1993J \citep{fransson1996,fransson1998}, but the X-rays were likely dominated by the thermal emission in this case \citep{suzuki1995,fransson1996,immler2001,nymark2009}. There is also a combined model for radio and X-ray properties of SN Ic 2002ap \citep{bjornsson2004}, but a quality of the observational data is not as great as that of SN 1993J to be conclusive. Recently, SN IIb 2011dh discovered soon after the explosion in M51\citep{griga2011,arcavi2011} provides a great data set from radio through X-rays \citep{soderberg2011,krauss2012,bietenholz2012}. 

SN 2011dh was discovered in M51 ($\sim 8.4$ Mpc) soon after the explosion. It was spectroscopically classified as Type IIb \citep{arcavi2011}, which initially shows strong Hydrogen lines but latter develops strong He lines \citep{filippenko1993,filippenko1997}. A progenitor star for Type IIb SNe (SNe) has been suggested to be a massive star which has lost most of its hydrogen envelope during its evolution, either by a strong wind or by a transfer to a binary companion \citep{nomoto1993,woosley1994}. The progenitors of SNe IIb are suggested to be diverse in the size of the hydrogen envelope, spanning from the `extended' progenitors of the red supergiant dimension (`eIIb' sub class, including SN 1993J) to the `compact' progenitors of the Wolf-Rayet star (`cIIb' sub class) \citep{chevalier2010}. Investigating high resolution images of SN 2011dh, a possible progenitor was reported as an yellow supergiant (YSG) with the radius $\sim 300 R_{\odot}$ \citep{maund2011,vandyk2011}. However, it has not been clarified yet if the detected star is the progenitor or a binary companion, or even an unrelated star. Indeed, SN 2011dh has been proposed to belong to the cIIb category based on the properties in the optical emission in the first few days \citep{arcavi2011}
and in the radio properties \citep{soderberg2011}. However, the theoretical interpretation of the early optical and radio emissions connecting to the size of the progenitor has not been completed yet, and a YSG progenitor may still be consistent with those observational constraints \citep{bersten2012}. 

In the optical wavelength, despite the difference in the initial phase lasting for a few days after the explosion, the subsequent evolution turned out to be very similar to SN 1993J \citep{arcavi2011,maund2011}. In this phase, the energy produced by the radioactive decay chain $^{56}$Ni $\to$ $^{56}$Co $\to$ $^{56}$Fe is reprocessed into the optical wavelength. The similarity in the light curve and spectra to SN 1993J suggests that the properties of the explosion of SN 2011dh, i.e., the energy and the ejecta mass, should be very similar to those of SN 1993J \citep{bersten2012}, despite the possible differences in the hydrogen envelope structure and the CSM environment. 

Thanks to its close distance and the very early discovery, intensive follow up observations have been performed not only in optical but also in radio and X-rays. In this paper, we analyze the radio and X-ray properties of SN 2011dh to constrain both the electron acceleration mechanism and the progenitor environment. The paper is organized as follow: 
In \S 2, we formulate the synchrotron radio emissions from the non-thermal electron populations and apply the model to SN 2011dh. In \S 3, we investigate the origin of the X-ray emission from SN 2011dh. The paper is closed in \S 4 with conclusions and discussion.

\section{Radio Emissions}

\subsection{Synchrotron Emission Model Description}

We follow the formalisms developed for the radio emissions from the interacting SNe by \citet{fransson1998,bjornsson2004} \citep[see also][]{chevalier1998,chevalier2006a,chevalier2006b}. An approximate expression for the optically thin synchrotron emission luminosity $\nu L_{\nu}$ is given as 
\begin{equation}
\nu L_{\nu} \sim \pi R^2 V n_{\rm e} \gamma_{\nu}^{2-p} m_{\rm e} c^2 
\left[1 + \frac{t_{\rm syn} (\nu)}{t} + 
\frac{t_{\rm syn} (\nu)}{t_{\rm other} (\nu)}\right]^{-1} \ .
\end{equation}
Here $R$ and $V$ are the position and the velocity of the forward shock, $n_{\rm e}$ is the number density of the relativistic electrons. For our `first' model, the energy distribution of the relativistic electrons is assumed to follow the power law with the index of $p$ extending down to $\gamma \sim 1$, following the standard prescription adopted in this kind of analysis. $t_{\rm syn}$ is the synchrotron cooling time scale. $t_{\rm other}$ is the time scale for other energy loss processes which do not emit at the radio frequency (e.g., inverse Compton (IC) scattering or Coulomb interaction). The Lorentz factor of the electrons emitting at frequency $\nu$ is 
$\gamma_{\nu} \sim 80 \nu_{10}^{0.5} B^{-0.5}$ (here $\nu_{10} = \nu/10^{10}$ Hz and $B$ is in gauss). 

Let us follow the equipartition argument for different energy channels, i.e., the standard prescription for the acceleration of the particles and the amplification of the magnetic fields. In this prescription, constant fractions ($\epsilon_{\rm e}$, $\epsilon_{B}$) of the shock-generated thermal energy are stored in the relativistic electrons and the magnetic field, and we have the following expressions: 
\begin{eqnarray}
B & \sim & 2.2 \times 10^{6} \epsilon_{B, -1}^{0.5} A_{\rm *}^{0.5} 
\left(\frac{V}{R}\right) {\rm gauss} \ ,\\
n_{\rm e} & \sim & 2.4 \times 10^{17} \left(\frac{p-2}{p-1}\right) \epsilon_{e, -1} A_{\rm *} 
\left(\frac{V}{R}\right)^2 {\rm cm}^{-3} \ . 
\end{eqnarray}
Hereafter $\epsilon_{\rm e, -1} \equiv \epsilon_{\rm e}/10^{-1}$ and $\epsilon_{B, -1} \equiv \epsilon_{B}/10^{-1}$. The CSM density scale is given by $A_{*}$, which is defined below and $A_{*} \sim 0.1 - 10$ for a WR or YSG progenitor. 
These two are connected as 
\begin{equation}
n_{\rm e} \sim 5 \times 10^{4} \alpha \left(\frac{p-2}{p-1}\right) B^{2} {\rm cm}^{-3} \ ,
\end{equation}
where $\alpha \equiv \epsilon_{\rm e}/\epsilon_{B}$ and $B$ is in gauss. 

The cooling time scales can be estimated as 
\begin{eqnarray}
t_{\rm synch} (\nu) & \sim & 110 \nu_{10}^{-0.5} B^{-1.5} \ {\rm days} \ , \\
t_{\rm IC} (\nu) & \sim & 1.7 \nu_{10}^{-0.5} B^{0.5} 
\left(\frac{L_{\rm SN}}{10^{42} {\rm erg s}^{-1}}\right)^{-1} \nonumber\\
& \times & \left(\frac{R}{10^{15} {\rm cm}}\right)^{2} {\rm days} \ , \\
t_{\rm Coulomb} (\nu) & \sim & 1200 \nu_{10}^{0.5} B^{-0.5} A_{*}^{-1} 
\left(\frac{R}{10^{15} {\rm cm}}\right)^{2} {\rm day} \ . 
\end{eqnarray}
Here $L_{\rm SN}$ is the bolometric luminosity of photospheric photons from the SN ejecta. Coulomb interactions can be important for the CSM of high density, e.g., in SNe IIp and extended SNe IIb (e.g., SN 1993J) \citep{fransson1998} with $A_{*} \gsim 100$. For SN 2011dh it is negligible, and we neglect the Coulomb loss in the following discussion.

So far no assumption has been made on the expansion of the shock wave. We assume that the CSM density is well represented by the steady state wind solution, 
\begin{equation}
\rho_{\rm CSM} = 5 \times 10^{11} A_{*} r^{-2} \ .
\end{equation}
The CSM density parameter $A_{*}$ is defined in a way so that $A_{*} \sim (\dot M / 10^{-5} M_{\odot} {\rm yr}^{-1}) (v_{\rm w} / 1,000 {\rm km s})^{-1}$ where $\dot M$ and $v_{\rm w}$ are the mass loss rate and wind velocity, respectively. A typical WR star has the mass loss properties of $\dot M \sim 10^{-5} M_{\odot}$ yr$^{-1}$ and $v_{\rm w} \sim 1,000$ km s$^{-1}$, thus $A_{*} \sim 1$. For a YSG case,  $\dot M \sim 10^{-6} M_{\odot}$ yr$^{-1}$ and $v_{\rm w} \sim 20$ km s$^{-1}$, thus $A_{*} \sim 5$. In each case, a variation in the value of $A_{*}$ spanning more than an order of magnitude is expected, thus it is difficult to discriminate these two progenitor scenarios solely from the value of $A_{*}$. 

The density distribution of the outermost SN ejecta is well described by a power law as a function of the velocity (with the index $n$). The hydrodynamic interaction between the materials both following the power law distributions is well approximated by a self-similar solution, and the propagation of the contact discontinuity is \citep{chevalier1982,chevalier2006b}, 
\begin{equation}
V_{\rm c} = 8 \times 10^9 E_{51}^{0.43} \left(\frac{M_{\rm SN}}{M_{\odot}}\right)^{-0.32} 
A_{\rm *}^{-0.12} t_{\rm d}^{-0.12} {\rm cm s}^{-1} \ . 
\end{equation}
The kinetic energy of the SN ejecta is in the unit of $10^{51}$ erg ($E_{51} \equiv E/10^{51}$ erg), and $t_{\rm d}$ is the time since the explosion in day. Here we adopt $n = 10.2$ which approximates the outer density structure of SNe IIb/Ib/Ic \citep{matzner1999,chevalier2008}. One has to take it in mind that this expression may contain an error in estimating the dynamic evolution especially since we take a specific value of $n$. Since the forward shock position is expected to be close to the contact discontinuity \citep{chevalier1982}, we assume $V = V_{\rm c}$ and $R = R_{\rm c}$ in the following analysis. 

For SN 2011dh, the CSM density is expected to be low \citep{soderberg2011}, and thus the synchrotron self-absorption (SSA) will be the dominant absorption process as in other SNe Ib/c and compact SNe IIb \citep{chevalier2006b}. The SSA frequency is 
\begin{equation}
\nu_{{\rm SSA}, 10} \sim 3 \times 10^{-5} \alpha^{2/7} R^{2/7} B^{9/7} \ .
\end{equation}

Now we have the complete set of equations to describe the radio emissions from SNe. The input parameters are $E_{51}$ and $M_{\rm ej}$ for the SN ejecta, $A_{*}$ for the CSM density, and $\epsilon_{\rm e}$ and $\epsilon_{B}$ for the microphysics at the shock wave. 

\subsection{Hydrodynamic Model Result}

We can further reduce the number of free parameters using various constraints. The SN ejecta properties are well constrained by the optical emission around the peak ($\sim 20$ days after the explosion):  The ejecta properties of SN 2011dh should be similar to those of SN 1993J (\S 1), i.e., $E_{51} \sim 1$ and $M_{\rm ej} \sim 2 - 4$ \citep{shigeyama1994,woosley1994}. Indeed, a detailed model for the optical emission of SN 2011dh suggests $E_{51} \sim 1$ and $M_{\rm ej} \sim 2 - 3 M_{\odot}$ \citep{bersten2012}. We hereafter take $E_{51} = 1$ and $M_{\rm ej} = 3 M_{\odot}$. For $L_{\rm SN}$ required to calculate the IC efficiency, we take a synthesized bolometric light curve from \citet{bersten2012} which best fits to the observed bolometric light curve. This is essentially identical to use the observed bolometric light curve \citep[e.g., Fig. 2 of ][]{arcavi2011} except for the IC cooling effect at $t_{\rm d} \lsim 5$ where the observed data are missing. Now, by specifying the three parameters ($A_{*}, \epsilon_{\rm e}, \epsilon_{B}$), we can compute the flux at any radio frequency at any epoch. 

Figure 1 shows one of the models that fit well the observed radio properties in multi bands. In this model, $A_{*} = 4$, $\epsilon_{\rm e} = 6 \times 10^{-3}$, and $\epsilon_{B} = 5 \times 10^{-2}$ (hereafter model A4). The effect of the IC cooling can be seen in the high frequency bands (16, 25, 36 GHz) around $t_{\rm d} \sim 20$, where the observation shows small suppression from the adiabatic model curves as is also seen in the model curves with the IC cooling. 

In addition to these three parameters $(A_{*}, \epsilon_{\rm e}, \epsilon_{B})$, the electron distribution index $p$ is a parameter, but we can derive this as $p \sim 3$ for the radio-emitting relativistic electrons ($\gamma \sim 50 -200$), independently from the other model parameters. This is the same as was taken in the previous analysis of SN 2011dh \citep{soderberg2011,krauss2012}, and also generally found in other SNe Ib/c and cIIb \citep{chevalier2006b}. The radio spectral evolution of SN 2011dh has been followed up to $\sim 100$ days when the cooling effect should be unimportant, and the photon spectral index there is consistent with $p = 3$. With this value, the temporal flux evolution in each band, as predicted to be $\propto t^{-1.36}$, is almost a perfect match to the observations as shown in Figure 1. Changing the value of $p$ is not acceptable, as in this case either the spectral index or the temporal index becomes inconsistent with the observations. As we do not model the details of the intermediate optical depth to SSA, the deviation in the peak in each band is an artifact -- in any case, the model is well constrained by the optical thick and thin phases simultaneously, and detailed modeling in the intermediate phase would not provide additional strong constraints. Except for this detail, the properties of the peaks in different bands are reproduced fairly well, supporting our assumption that the SSA is the dominant absorption process in the radio wavelength. 

\subsection{Efficiency of the electron acceleration}

In our analysis, $A_{*}, \epsilon_{\rm e}$, and $\epsilon_{B}$ are not mutually independent, unlike in the previous studies \citep{soderberg2011,krauss2012}. This is because we construct the model taking into account the optical emission properties, using the SN properties and resulting shock wave evolution independently from the radio study. Our model requires a rather small value of $\epsilon_{\rm e}$ as compared to the value typically assumed in the analysis of the radio properties of SNe (i.e., $\epsilon_{\rm e} = \epsilon_{B} = 0.1$). A question is what to extent the degeneracy is involved in the model parameters to produce similar light curves and spectra. From the SSA peak, we constrain the combination of $A_{*} \epsilon_{\rm e}^{8/19} \epsilon_{B}^{11/19}$ \citep{chevalier1998} and from the optically thin flux we constrain $A_{*}^{1.64} \epsilon_{\rm e} \epsilon_{B}$. The latter scaling relation is obtained by substituting equations 2 -- 5 into 
equation 1, for the optically thin and adiabatic phase assuming $p=3$. Thus, $\epsilon_{\rm e}$ and $\epsilon_{B}$ can separately be obtained for given value of $A_{*}$ as long as the cooling processes are not important in the wavelength of interest. In Figure 2, we show the combination of  $(\epsilon_{\rm e}, \epsilon_{B})$ which produces essentially identical light curves to fit the observed light curves. The required value for $\epsilon_{\rm e}$ is rather insensitive to $A_{*}$ as is evident from the scaling relations as described above. As an illustration, in Figure 3 we show model A with $A_{*} = 30$ with $\epsilon_{\rm e}$ and $\epsilon_B$ following these relations (model A30). The radio light curves are similar to those of model A4 (the effect of increasing $A_{*}$ will be further discussed below).

From Figure 2, there is another important information. Since $\epsilon_{B}$ cannot exceed $\sim 1/3$, we set a strong constraint as $A_{*} \gsim 2$. With this constraint, if we further assume that $A_{*} \sim 2 - 10$ as expected for SN 2011dh (either a WR or YSG progenitor), then we constrain that $\epsilon_{\rm e}$ must be in the range between $5 \times 10^{-3}$ and $0.01$, $\epsilon_{B} \gsim 0.01$, and  $\alpha$ must be below unity ($0.02 - 1$). 

The difference in our model and the interpretation in the previous studies \citep{soderberg2011,krauss2012} should be clarified. Using the scaling relation for the SSA peaks, \citet{soderberg2011} estimated $A_{*} \sim 3$ {\em assuming} $\epsilon_{\rm e} = \epsilon_{B} = 0.1$. They also provided the estimate with $\epsilon_{\rm e} = 0.3$, $\epsilon_{B} = 0.01$ and $A_{*} = 6$ to explain the X-ray luminosity by the inverse Compton mechanism (see \S 3 for details on the X-ray properties). The similar values were obtained in another paper by the same group \citep{krauss2012}. In our models, these sets of parameters never reproduce the observed radio light curves. The reason for this difference is simple: we do not allow the dynamics of the interacting region as a free parameter but rather use the one consistent with the optical observations. Figure 4 shows the evolution of the shock wave radius in our model and those estimated previously \citep{krauss2012} as well as the constraints from the VLBI measurements \citep{bietenholz2012}. It is seen that the shock radius estimated previously is smaller than in our model. If we fit the evolution of radius estimated by \citet{krauss2012}, we need to increase the ejecta mass to $\gsim 10 M_{\odot}$, which is clearly rejected from the optical properties (\S 1). This argument relies on the applicability of the self-similar solution for the shock dynamics (equation 9) which can be dependent on the slope of the outermost ejecta. While we have followed the widely accepted recipe used in many literatures \citep[e.g.,][]{chevalier2006b,chevalier2008}, it is possible that this description may contain an error. Still, we believe the difference between our result and the solution for $\epsilon_{\rm e} = \epsilon_{B} = 0.1$ is too large to be explained by the possible error in the self-similar dynamic solution. Our result here suggests that one must be careful to convert the properties of the SSA peaks in radio wavelengths to the real physical size if one wants to do it with the accuracy better than a factor of two; if one adopts values for the microphysics parameters ($\epsilon_{\rm e}, \epsilon_{B}$) {\em a priori}, it does not provide a check in the internal consistency between the dynamic evolution and the microphysics parameters. The conventionally assumed values, $\epsilon_{\rm e} = \epsilon_{B} = 0.1$, do not provide a solution consistent both in radio and optical according to our analysis. 

In this sense, the VLBI measurement is very important since it is independent from the microphysics parameters \citep{chevalier1998,bietenholz2012}, and this should be used as an independent and strong constraint. Although the radius predicted in our model is large, it is still within the standard error of the VLBI measurements. Indeed, the VLBI measurements provide another independent constraint on the CSM density: $A_{*} \gsim 4$ if we take $E_{51} = 1$ and $M_{\rm ej} = 3 M_{\odot}$. 

For a demonstration purpose, we show the synthetic radio light curves for the `low velocity' model B in Figure 5. In Model B, we artificially changed the ejecta properties so that the radio light curves are fit with $\epsilon_{\rm e} = \epsilon_{B} = 0.1$. In terms of the self-similar solution, we require $M_{\rm ej} = 10.5 M_{\odot}$ if $E_{51} = 1$, and $A_{*} = 0.9$ in model B (hereafter model B1). Note that there is no degeneracy in $M_{\rm ej}$ and $A_{*}$ if we fix the other three parameters. 
The radius is smaller than in model A4, but still larger than the previous SSA estimate (note that we do not have to obtain the same parameter set, since the coupled dynamics and the emission processes provide constraints that are not always satisfied by the simple SSA estimate). We can obtain the evolution of the radius identical to that given by the SSA estimate, but it requires the large value of $A_{*}$ and a different set of ($\epsilon_{\rm e}$ and $\epsilon_{B}$) as shown in Figure 2. It is seen that model B1 predicts the radio light curves similar to model A4 as shown in Figure 5. The main difference is that the inverse Compton cooling is more important in model B1 due to the smaller radius and larger photon density, but the difference is still not large to discriminate model A4 and model B1 purely from the radio light curves. If we take model B for the evolution of the shock radius, then the constraint is placed on the CSM density as $0.5 \lsim A_{*} \lsim 30$ (Fig. 2).

\section{X-Ray Emission}

SN 2011dh turned out to be a strong X-ray emitter \citep{soderberg2011} (Fig. 6). On average the flux evolution in the 0.3 - 8 keV range followed the dependence of $\propto t^{-1}$, but it showed a complicated evolution in detail. There is a hint of the initial fast decay from $\sim 3 \times 10^{39}$ erg s$^{-1}$ at 4 days to $\sim 10^{39}$ erg s$^{-1}$ at 6 days. It was then followed by almost a flat evolution, reaching to $\sim 2 \times 10^{39}$ erg s$^{-1}$ at $\sim 10$ days. Around 20 days it started a rather fast decline, from $\sim 10^{39}$ erg s$^{-1}$ to $\sim 2 \times 10^{38}$ erg s$^{-1}$ in about 20 days (i.e., $\propto t^{-2}$ or even steeper). 

\subsection{Difficulties in thermal and synchrotron emissions}

There are three mechanisms suggested so far for the X-ray emission from SNe Ib/c and compact SNe IIb, i.e., thermal emission, synchrotron, and inverse Compton \citep[e.g.,][]{chevalier2006a,chevalier2006b}. For SN 2011dh, \citet{soderberg2011} discussed these possibilities, and preferred the inverse Compton scenario. For the thermal emission, the free-free emission from the reverse shock likely dominates in the 0.3 - 8 keV range, and with $A_{*} = 4$ the predicted X-ray luminosity is $\sim 4 \times 10^{37}$ erg s$^{-1}$, nearly two orders of magnitudes fainter than observed. The free-free X-ray luminosity can be comparable to the observed luminosity only if $A_{*} \sim 30$. Such a high density CSM is not rejected from the radio properties alone (Fig. 3). The IC cooling is important, but it is still consistent with the observed light curves (see below for further discussion on the IC cooling effect). Also, because of the high shock velocity, the free-free absorption time scale is estimated to be a few days \citep{chevalier2006a}, i.e., still negligible compared to the SSA and thus consistent with the observed radio properties. 

However, following arguments make this interpretation disfavored: (1) the mass loss rate is then an order of magnitude larger than expected from the typical mass loss rate of a WR or YSG star. (2) Related to this argument, assuming the microphysics parameters ($\epsilon_{\rm e}$ and $\epsilon_{B}$) are shared by other SNe Ib/c, then it will lead to the increase of the CSM density (i.e., $A_{*}$) by a factor of $\sim 30$ {\em in other SNe Ib/c} as well -- although the high CSM may apply to SNe IIb and in particular SN 20011dh with the possible YSG progenitor (which may require a special binary channel as a progenitor scenario), it is hard to reconcile such a high density CSM with the WR progenitor generally accepted for SNe Ib/c. (3) The X-ray light curve shows a complicated temporal structure, while the free-free emission predicts a simple power law decay unless the CSM structure deviates significantly from a smooth steady-wind solution. Such a variation in the CSM structure should be seen as a modulation in the radio light curves \citep{ryder2004,soderberg2006,wellons2012}, but it is not seen in SN 2011dh. 

The synchrotron emission is even less likely. Phenomenologically, if we extrapolate the radio spectrum to the X-ray band then $\nu L_{\nu}$ (X-ray) $\sim \nu L_{\nu}$ (radio) $\sim 10^{37}$ erg s$^{-1}$ \citep{soderberg2011}. The situation is even worse if we consider the cooling processes. At 10 days, the synchrotron cooling frequency is at most $\lsim 10^{13}$ Hz. The Compton cooling frequency is less  model dependent, and it must be below $\sim 10^{11}$ Hz even if a very strong $B \sim 10$ gauss is assumed. Thus, the inverse Compton cooling is important to reduce the number of high energy relativistic electrons emitting in the X-ray frequency. With this effect taken into account, the synchrotron X-ray luminosity must be many order of magnitudes smaller than observed. Although X-ray emission from some SNe Ib/c (typically those in a later phase than considered here) could be attributed to the synchrotron emission if the intrinsic relativistic electron spectrum flattens at the energy higher than probed by radio emissions (e.g., at $\gamma \sim 1000$) \citep{chevalier2006b}, this mechanism would not operate for the early X-ray emission at $\sim 10$ days as observed for SN 2011dh. The reason is evident from the above argument about the cooling processes. Indeed, the Compton cooling energy of the electrons is rather insensitive to any underlying model assumption (i.e., it depends on the SN luminosity and the shock radius, but independent from the microphysics parameters), and it is $\gamma \sim 150-200$ for SN 2011dh at 10 days. Thus, even if the spectral flattening appears in the higher energy, because of the cooling the expected X-ray emission would not exceed the value estimated with the extrapolation of the radio spectrum. 

\subsection{Inverse Compton mechanism}

The remaining, and most attractive, possibility is the IC scattering of the SN photospheric photons by the relativistic electrons. 
We can estimate the inverse Compton luminosity by a formula similar to the synchrotron radio emission: 
\begin{equation}
\nu L_{\nu} \sim \pi R^{2} V n_{\rm e} \gamma^{2-p} mec^2 \left[1+\frac{t_{\rm d}}{t_{\rm ic} (\gamma)}\right] \ , 
\end{equation}
The SN photospheric emission has the characteristic energy of $\sim 1$ eV, thus the electrons with $\gamma \sim \sqrt{1 {\rm keV}/1{\rm eV}} \sim 30$ up-scatter the photospheric photons to $\sim 1$ keV. On the other hand, the characteristic energy of the electrons corresponding to the IC cooling frequency is not sensitive to the model details, and it is $\gamma_{\rm IC} \sim 150 - 200$ at 10 days. Thus the electrons emitting the X-rays through the IC mechanism is in the adiabatic phase. 

In this paper we show that the IC is indeed a favored interpretation. Although it has been also favored by \citet{soderberg2011}, our conclusion is based on a different line of arguments than theirs. Figure 6 shows the X-ray light curves from Models A4 and B1, both fit the radio light curves fairly well. The predicted luminosity is about 25 to 100 times fainter than observed, as is consistent with the analysis by \citet{soderberg2011}. However, given the striking similarity of the IC prediction to the observed qualitative behavior as well as the difficulties in the other mechanisms, we investigate if there is a solution to remedy the problem and provide a unified explanation for the radio and X-ray properties. 

It has been suggested that changing the microscopic parameters, assuming effective electron acceleration ($\epsilon_{\rm e} \sim 0.3$) and inefficient magnetic field generation ($\epsilon_{B} \sim 0.01$) will do to fit the X-ray data by the IC \citep{soderberg2011}. However, this option is rejected from our analysis, which constrains these parameters to the opposite direction (Fig. 2). 
A second argument is from a consideration on the electron number density. Assuming the standard single power law distribution, at 10 days the number density of relativistic electrons assumed in this formalism is $n_{\rm rel} \sim 6.4 \times 10^{5} (\epsilon_{\rm e}/0.1) (A_{*}/4)$, while the total electron number density is $n_{\rm e} \sim 1.3 \times 10^{5} (A_{*}/4)$ (including thermal electrons). Therefore, the condition $\epsilon_{\rm e} \lsim 0.02$ must be met, otherwise the model is self-inconsistent. 

Finally, even if we ignore these discrepancies and allow a large value of $\epsilon_{\rm e}$, such a model is in principal not accommodated with the radio properties. For the IC scattering works effectively, the low value of $B \lsim 0.5$ Gauss is required. Then, since the energy of the electron suffering the significant IC cooling is model-insensitive and $\gamma_{\rm IC} \sim 150 - 200$ during  $10 -30$ days, this inevitably means that the radio emission above $\nu \gsim 20$ GHz is significantly suppressed by the IC cooling effect. This is not seen in the radio data -- the observed radio spectral index does not show the expected considerable steepening at the high frequency (note that there is a hint of the small IC cooling effect there, but this behavior is compatible to that predicted by our standard model A4: Fig. 1). 

To highlight the final point, in Figure 7 we show an example of the radio light curves for a model in which the IC cooling is effective enough to explain the X-ray luminosity (Fig. 6). Since model A does not allow a large value of $\epsilon_{\rm e}$, we start with model B and change $A_{*}$ so that we reaches the X-ray luminosity comparable to the observations, while fixing the microphysics parameters so that the radio emission {\em without} the IC cooling fits the radio light curves (i.e., along the constraint on $\epsilon_{\rm e}$ and $\epsilon_{B}$ in Fig. 2). Figure 7 shows the model radio light curves. We require $A_{*} \sim 20$ (i.e., model B20). This model predicts the X-ray light curve through the IC mechanism fairly well (Fig. 6), but then the IC cooling effect becomes significant and alters the spectral index of the radio emission at $\gsim 20$ GHz. It never fits to the observed radio properties. This is not a model-specific problem, but a general problem in the models with large value of $\epsilon_{\rm e}$ since such a model inevitably requires small $B$ and the resulting radio emitting electrons' energy above the IC cooling energy. 

\subsection{What is missing? -- the pre-accelerated electron population}

As shown in \S 3.2, the IC scattering cannot explain the X-ray luminosity in a self-consistent manner. However, given the failure of the synchrotron mechanism and the required high value of $A_{*}$ in the thermal scenario, and also the striking similarity of the IC X-ray model curve evolutioin to the observed one, we further investigate what is required if the X-ray is indeed a consequence of the IC mechanism. 

All the three problems in the IC scenario found in \S 3.2 are indeed related to one single assumption in our (and conventional) model prescription - the single power law distribution of the relativistic electrons. Alternatively, we suggest that this assumption is not appropriate in the low energy regime, and that the discrepancy in the predicted and observed X-ray flux does reflect the deviation of the electron distribution from this standard/conventional formalism. The strong constraint from the radio properties is obtained only for the electrons with $\gamma \sim 50 - 200$. On the other hand, the electrons responsible for the X-ray emission through the IC mechanism have the lower energy, about $\gamma \sim 30$. Thus, a consistent view on the radio and X-ray properties of SN 2011dh can only be obtained if the electron number density at $\gamma \sim 30$ is by about two orders of magnitudes larger than the extrapolation from that at $\gamma \sim 50 - 200$. Adopting this distribution, the X-ray light curve is well reproduced (Fig. 6). The energy range of this additional electron population is below the energy probed by the radio properties, thus the resulting radio light curve is hardly affected and is consistent with that derived assuming the single power law distribution. This relatively low energy electron population must be peaked around this energy regime ($\gamma \sim 20 - 30$) and cannot extend down to $\gamma \sim 1$, otherwise it violates the electron number conservation. The derived distribution is shown in Figure 8. We suggest this population represents the electrons accelerated by the `pre-acceleration' mechanism. With this distribution of relativistic electrons, the radio (Fig. 1) and X-ray (Fig. 6) emissions are simultaneously explained.

\section{Discussion and Conclusions}

In this paper, we have studied the radio and X-ray properties of SN IIb 2011dh. From the radio light curves in multi-bands, we have obtained strong constraints on the efficiency of the electron acceleration ($\epsilon_{\rm e}$) and the magnetic field generation/amplification ($\epsilon_{B}$) behind the shock wave. 

The model developed here is only weakly constraining the CSM environment. From the requirement that $\epsilon_{B} \lsim 1/3$ and the shock radius through the VLBI measurement, we have a strong constraint as $A_{*} \gsim 4$. The fact that we see only a slight effect of the IC cooling in the radio light curves places a constraint that $\alpha$ cannot be much larger than unity: In the model B sequence $A_{*} \sim 20$ is clearly rejected, while in the model A sequence $A_{*} \sim 30$ is still acceptable. The allowed range is consistent with both the WR and YSG progenitor scenarios. If $A_{*} \sim 30$, the X-ray luminosity can be explained by the thermal emission from the materials behind the (reverse) shock wave although the fit is rather poor to explain the time evolution. 
Although the required CSM density exceeds the typical mass loss rate of a WR or YSG star by an order of magnitude, a variant of a binary interaction scenario to produce the SN IIb progenitor may (although does not have to) lead to such a high mass loss rate in the end of the evolution \citep{benvenuto2012}. The radio data alone do not reject this possibility for a particular case of SN 2011dh. However, if we naturally extend our argument that the microphysics of the magnetic field amplification and the particle acceleration is similar between SN 2011dh and other SNe Ib/c, then the required CSM density for other SNe Ib/c should increase by a factor of $\sim 30$ and never fit to the Galactic WR mass loss rate. Thus, we have investigated another possibility -- how to explain the radio and X-ray properties simultaneously for the standard mass loss rate of $A_{*} \sim 1 - 10$.

The IC cooling effect is important in constraining the properties of the microphysics and the mass loss (as mentioned above for the upper limit of $A_{*}$). The large values of $A_{*}$ and resulting  large $\epsilon_{\rm e}$ are rejected because such a situation will predict the significant IC cooling in the {\em radio} wavelength. We thus obtained that $\epsilon_{\rm e} \lsim 0.01$, much lower than conventionally assumed for the SN-CSM interaction. Such a low value of $\epsilon_{\rm e}$ was also derived for SN eIIb 1993J \citep{fransson1998}, and also not rejected for SNe IIp in general \citep{chevalier2006a}. Thus, it can be a property shared by SNe Ib/c and cIIb as well. Since $\epsilon_{B}$ can as large as conventionally assumed (i.e., $\epsilon_{B} = 0.05$ in our standard model A4), the mass loss rate estimated through the SSA scaling \citep[e.g., ][]{chevalier2006b} will not need a drastic revision. 

This argument on the IC cooling effect led us to conclude that the X-ray properties of SN 2011dh cannot be accommodated by the IC mechanism, as well as the thermal and synchrotron, {\em as long as the single power law distribution of the relativistic electrons is assumed}. Alternatively, we suggest that the single power law distribution is not appropriate for the energy range probed by radio and X-rays from SN 2011dh (i.e., $\gamma \lsim 200$), and that the discrepancy does reflect the deviation of the electron distribution from the conventionally assumed power law. To explain the radio and X-ray properties simultaneously, we need a population of relativistic electrons peaking around $\gamma \sim 20 - 30$ in addition to the power law component extending to the higher energy (in which those with $\gamma \sim 50 - 200$ are probed by the radio observations). This low-energy population indeed occupies a major part of the relativistic electrons both in number and energy. We suggest this electron population represent the electrons accelerated by the `pre'-acceleration mechanism, and the power law component represents ones further accelerated by the DSA-like mechanism. Although the argument is based on modeling radio and X-ray properties of SN 2011dh, we speculate that these properties may be shared by other SNe as well, given that SN 2011dh is a canonical SN in its radio properties and that the energy range of electrons probed in the radio and X-rays in other SNe cIIb/Ib/Ic should be similar to the case of SN 2011dh. Good X-ray data are generally missing in other SNe IIb/Ib/Ic, and the future intensive observations in radio and X-ray will be valuable to study the low energy relativistic electron population as we suggest in this paper.

The typical energy of the pre-accelerated electrons we have found is $\gamma \sim 20 - 30$, thus the number density of the relativistic electrons at 10 days is $n_{\rm e} \sim 10^{4}$ cm$^{-3}$. This requires $\sim 1 - 10$\% of the thermal electrons suffer from the pre-acceleration, rather than $\sim 100$\% of the electrons accelerated to the relativistic energy required when assuming the single power law electron distribution. Since this reduction in the number density is compensated by the increase of the energy content of each electron by the typical energy $\gamma \sim 20 - 30$, the `intrinsic' efficiency of the acceleration including this pre-acceleration population is similar to the `apparent' efficiency of $\epsilon_{\rm e} \lsim 0.01$ derived by the radio modeling alone. 

Our finding provides an observational constraint on the nature of the pre-acceleration mechanism. The typical energy scale, $\gamma \sim 20 - 30$, suggests that the pre-acceleration of the electron takes place simultaneously and coupled with the amplification of the magnetic field \citep{frederiksen2003,hededal2004}. The process would equalize the kinetic energy of a thermal proton and a (pre-)accelerated electron \citep{frederiksen2003}, thus $\gamma \sim (m_{\rm p}/m_{\rm e}) (V/c)^{2} \sim 20$ (where $c$ is the speed of light, and $V\sim 0.1 c$ is the velocity of the shock wave). This is exactly the energy scale we need for SN 2011dh. Then, about 1\% of these pre-accelerated electrons are further accelerated through the DSA-like process, as evidenced by a power law behavior of the electron distribution in the higher energy as derived by the radio properties.

\acknowledgements 
This research is supported by World Premier International Research Center
Initiative (WPI Initiative), MEXT, Japan. K. M. acknowledges financial support by Grant-in-Aid for Scientific Research for young Scientists (23740141). 
K.M. thanks Melina Bersten for discussion on the hydrodynamic and progenitor properties of SN 2011dh, Takashi Moriya for general discussion, and Ken'ichi Nomoto for continuous encouragement.


\clearpage
\begin{figure*}
\begin{center}
        \begin{minipage}[]{0.8\textwidth}
                \epsscale{1.0}
                \plotone{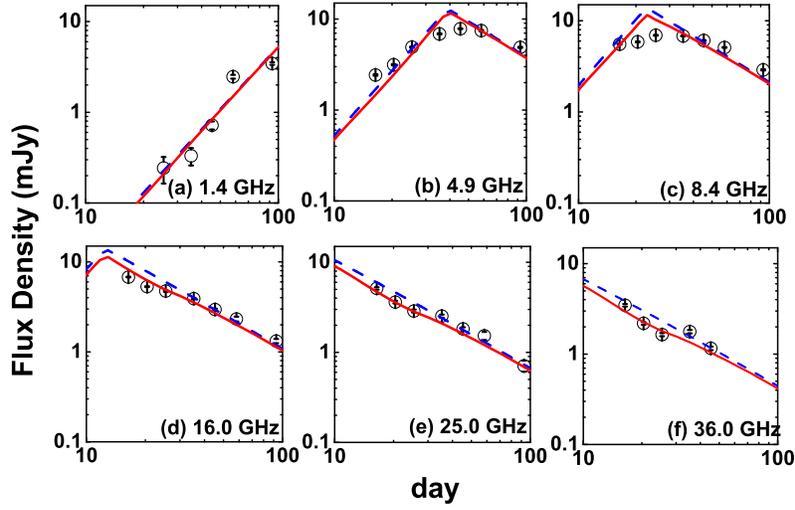}
        \end{minipage}
\end{center}
\caption
{Multi frequency radio light curves of Model A4 (red solid) as compared with those of SN 2011dh. The parameters are 
$(A_{*}, \epsilon_{\rm e}, \epsilon_{B}) = (4, 6 \times 10^{-3}, 5 \times 10^{-2})$. 
The synthetic light curves without the inverse Compton cooling are also shown (blue dashed). 
\label{fig_radioLC_modelA4}}
\end{figure*}

\begin{figure*}
\begin{center}
        \begin{minipage}[]{0.6\textwidth}
                \epsscale{1.0}
                \plotone{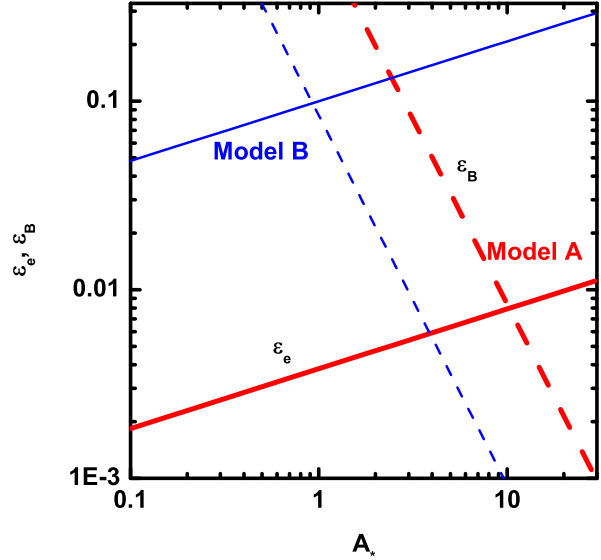}
        \end{minipage}
\end{center}
\caption
{The values of $\epsilon_{\rm e}$ and $\epsilon_{B}$ derived for SN 2011dh through the radio light curves, as a function of $A_{*}$. 
\label{fig_equipartition}}
\end{figure*}

\clearpage
\begin{figure*}
\begin{center}
        \begin{minipage}[]{0.8\textwidth}
                \epsscale{1.0}
                \plotone{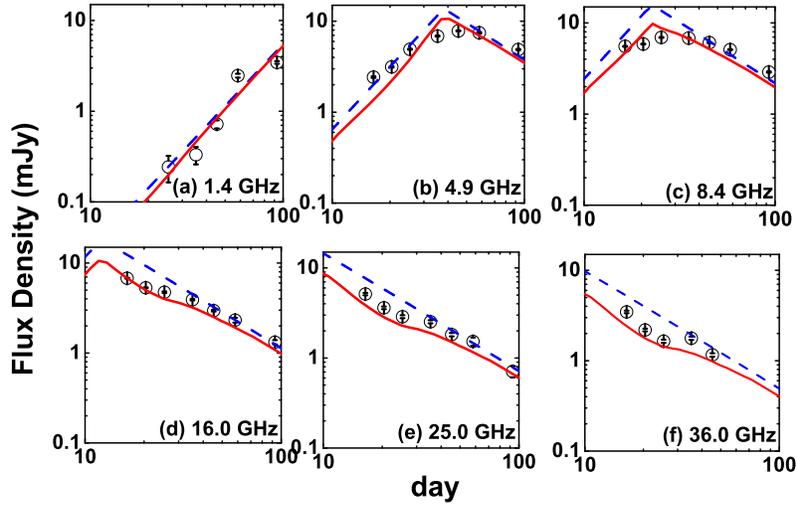}
        \end{minipage}
\end{center}
\caption
{Multi frequency radio light curves of Model A30 (red solid) as compared with those of SN 2011dh. The parameters are $(A_{*}, \epsilon_{\rm e}, \epsilon_{B}) = (30, 0.01, 9.7 \times 10^{-4})$ (Fig. 2). The synthetic light curves without the inverse Compton cooling are also shown (blue dashed). 
\label{fig_radioLC_modelA30}}
\end{figure*}

\begin{figure*}
\begin{center}
        \begin{minipage}[]{0.6\textwidth}
                \epsscale{1.0}
                \plotone{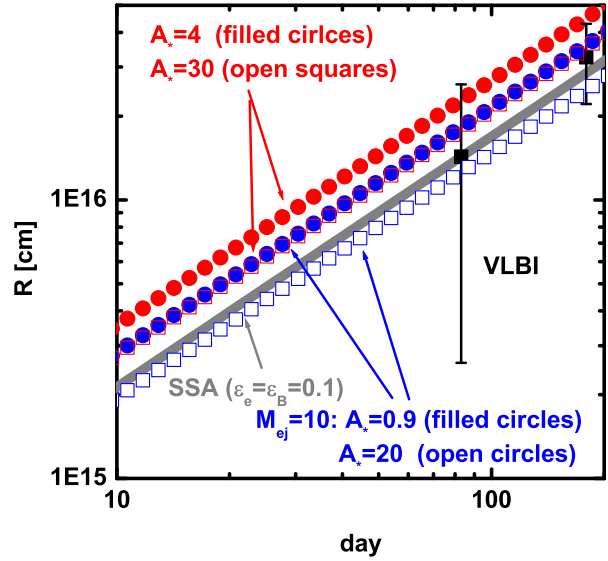}
        \end{minipage}
\end{center}
\caption
{The evolution of the radius of the interaction region. Thick gray line is that estimated with the scaling relation for the SSA peaks \citep{chevalier1998} assuming $\epsilon_{\rm e} = \epsilon_{B} = 0.1$\citep{soderberg2011,krauss2012}. The VLBI measurements are given by filled squares\citep{bietenholz2012}. Model A is shown by red filled circles (A4: $A_{*} = 4$) and by red open squares (A30: $A_{*} = 30$), while model B is by blue filled circles (B1: $A_{*} = 0.9$) and blue open squares (B20: $A_{*} = 20$).  
\label{fig_radius}}
\end{figure*}

\clearpage
\begin{figure*}
\begin{center}
        \begin{minipage}[]{0.8\textwidth}
                \epsscale{1.0}
                \plotone{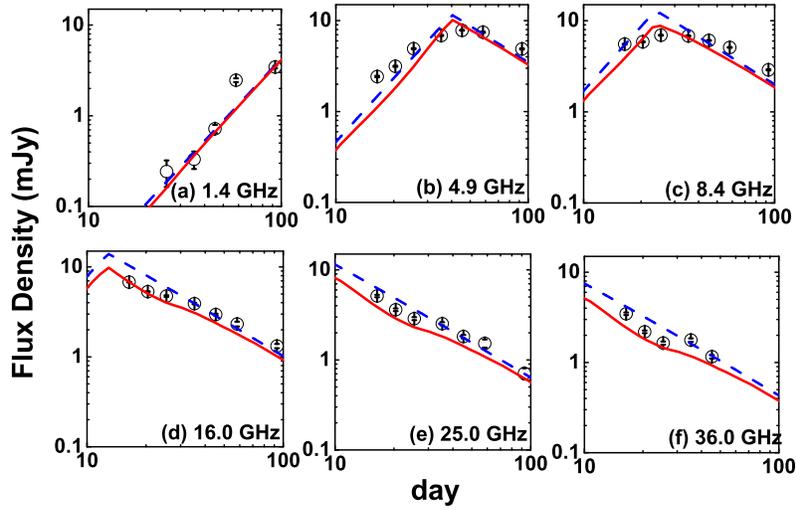}
        \end{minipage}
\end{center}
\caption
{Multi frequency radio light curves of Model B1 (red solid) as compared with those of SN 2011dh. The parameters are $(A_{*}, \epsilon_{\rm e}, \epsilon_{B}) = (0.9, 0.1, 0.1)$. The synthetic light curves without the inverse Compton cooling are also shown (blue dashed). 
\label{fig_radioLC_modelB1}}
\end{figure*}

\begin{figure*}
\begin{center}
        \begin{minipage}[]{0.6\textwidth}
                \epsscale{1.0}
                \plotone{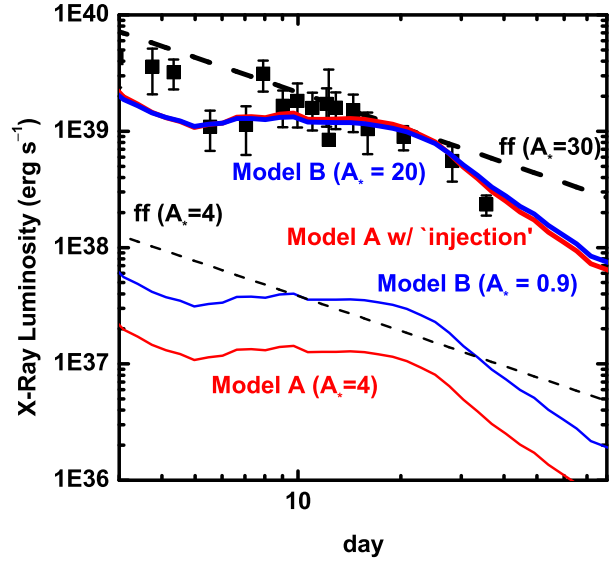}
        \end{minipage}
\end{center}
\caption
{X-ray light curves of SN 2011dh (0.1 - 8 keV) as compared with some models discussed in the text. Dashed curves are for the free-free emission, with $A_{*} = 4$ (thin) and $A_{*} = 30$ (thick). Thin lines are for the inverse Compton scattering, for model A4 ($A_{*} = 4$) and model B1 ($A_{*} = 0.9$). The inverse Compton emission in model B20 ($A_{*} = 20$) is shown by the thick blue line. Finally, model A4 but introducing another electron population peaking at $\gamma \sim 25$ is shown by thick red line. 
\label{fig_xLC}}
\end{figure*}

\clearpage
\begin{figure*}
\begin{center}
        \begin{minipage}[]{0.8\textwidth}
                \epsscale{1.0}
                \plotone{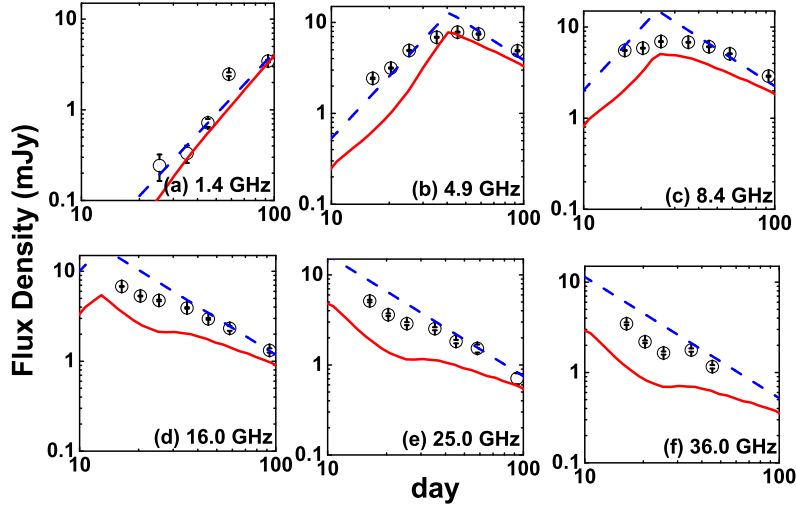}
        \end{minipage}
\end{center}
\caption
{Multi frequency radio light curves of Model B20 (red solid) with large $\epsilon_{\rm e}$ to reproduce the X-ray light curve by the inverse Compton scattering. The parameters are $(A_{*}, \epsilon_{\rm e}, \epsilon_{B}) = (20, 0.26, 2.5 \times 10^{-4})$. The synthetic light curves without the inverse Compton cooling are also shown (blue dashed). The effect of the inverse Compton cooling is important in the radio wavelengths. 
\label{fig_radioLC_modelB20}}
\end{figure*}

\begin{figure*}
\begin{center}
        \begin{minipage}[]{0.6\textwidth}
                \epsscale{1.0}
                \plotone{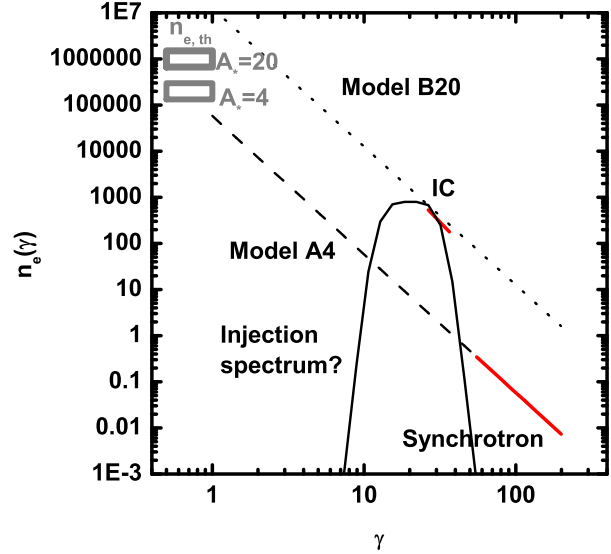}
        \end{minipage}
\end{center}
\caption
{Constraints on the energy distribution of relativistic electrons at the shock front of SN 2011dh. The scale is for 10 days since the explosion. The dashed line is a standard single power law in model A4 to fit the optically thin radio synchrotron emission. Along the line, the red region is the constraint from the radio light curves in the optically thin phase. The red region with the label `IC' is the electron population required to explain the X-ray emission at $\sim 1$ keV by the IC mechanism. For comparison, we show the electron spectrum in model B with $A_{*} = 20$ (B20) that explain the X-ray emission by the IC (but failed to explain the radio properties). This model B line is similar to the `IC' population derived based on model A, demonstrating that the argument is rather model-independent. The gray box at $\gamma \sim 1$ shows the thermal electron population for an appropriate range of the CSM density and the shocked radius. Finally, a schematic picture is shown for the distribution required for the low energy electrons to explain both the radio (by synchrotron) and the X-ray (by the IC), labeled as the `injection spectrum?'. 
\label{fig_electronspectrum}}
\end{figure*}

\end{document}